\documentclass{aip-cp}

\usepackage[numbers]{natbib}
\usepackage{graphicx}
\usepackage{url}

\begin{document}

\title{Theory Considerations For Nucleosynthesis Beyond Fe With
Special Emphasis On p-Nuclei In Massive Stars}

\author[aff1,aff2,aff3]{T. Rauscher\corref{cor1}}

\author[aff4,aff3]{N. Nishimura}

\author[aff4,aff3,aff5]{R. Hirschi}

\affil[aff1]{Centre for Astrophysics Research, University of Hertfordshire, Hatfield AL10 9AB, United Kingdom.}
\affil[aff2]{Department of Physics, University of Basel, 4052 Basel, Switzerland.}
\affil[aff3]{UK Network for Bridging Disciplines of Galactic Chemical Evolution (BRIDGCE), \url{http://www.astro.keele.ac.uk/bridgce}, United Kingdom.}
\affil[aff4]{Astrophysics group, Faculty of Natural Sciences, Keele University, Keele ST5~5BG, UK.}
\affil[aff5]{Kavli IPMU, The University of Tokyo, Kashiwa 277--8583, Japan}
\corresp[cor1]{Corresponding author: t.rauscher@herts.ac.uk}

\maketitle

\begin{abstract}
Nucleosynthesis of heavy elements requires the use of different experimental and theoretical methods to determine astrophysical reaction rates than light element nucleosynthesis. Additionally, there are also larger uncertainties involved in the astrophysical models, both because the sites are not well known and because of differing numerical treatments in different models. As an example for the latter, the production of p-nuclei is compared in two different stellar models, demonstrating that a model widely used  for postproduction calculations may have a zone grid too coarse to follow the synthesis of p-nuclei in detail.
\end{abstract}

\section{INTRODUCTION}
Investigations of nucleosynthesis involving intermediate and heavy nuclei often have to proceed in a different manner than those in light element nucleosynthesis. On one hand, with the exception of the s-process, production sites are less well established. Some of the underlying processes are less well constrained and further processes and/or various sites may additionally contribute. Reaction networks are large and postprocessing apporaches have to be used. On the other hand, the nuclear physics properties required to obtain astropysical reaction rates as input to reaction network calculations, and the relation between experiment and theory, are different for light nuclei than for intermediate and heavy species.
Special for nucleosynthesis beyond Fe are \cite{advances}
\begin{enumerate}
\item generally higher plasma temperatures,
\item higher matter density during the astrophysical burning process,
\item higher intrinsic nuclear level density of the involved intermediate and heavy nuclei, and
\item higher Coulomb barriers.
\end{enumerate}
Items 1 and 4 result in higher effective interaction energies than in light element nucleosynthesis but still low by nuclear physics standards, ranging from a few keV to a couple hundred keV for neutron captures in s- and r-process. In charged particle reactions the effective energies are shifted to a few MeV, depending on the specific process and the charge of the nucleus and projectile (or ejectile) \cite{energywindows}. In the production of p-nuclei (see below), reactions dominated by proton widths have their relevant energy range at a few MeV and such dominated by $\alpha$ widths up to $\approx 9-10$ MeV.
Points 1 and 2 also imply that unstable nuclides are involved, which can be studied only in a limited manner in the laboratory, if at all. The higher inherent nuclear level density of heavier nuclei combined with the higher temperature leads -- items 1 and 3 -- to considerable contributions of transitions from excited states of the target nuclei (see next section).

Generally speaking, these circumstances conspire to simplify a theoretical treatment (with exception of reactions on magic nuclei and close to the driplines, see, e.g., \cite{raureview}) and complicate an experimental constraint of \textit{stellar} rates. The many transitions (from target states to final states, mostly via compound states) involved in heavy element nucleosynthesis lend themselves to the use of averaged quantities (as in the Hauser-Feshbach model \cite{haufesh,raureview}) in their prediction (for further important reaction mechanisms and their sensitivity to nuclear properties, see \cite{raureview,sensis}. Contrary to what is observed for reactions with light nuclei and/or higher interaction energies, the sensitivity of astrophysically relevant reaction cross sections of intermediate and heavy nuclei is determined by the charged particle widths instead of the $\gamma$ width. This is because the charged particle widths become smaller than the $\gamma$ widths due to the higher Coulomb barriers. This simplifies the treatment as photon widths are difficult to predict at low $\gamma$ energies. On the other hand, the large number of transitions restricts the applicability of direct and indirect experimental approaches studying a few transitions, as usually applied in the study of light nuclei, even when dealing with stable nuclides. Finally, only reactions on targets in the ground state (g.s.) are accessible in the laboratory whereas heavy element nucleosynthesis not only involves highly unstable targets but also such in excited states.

The above facts and possible problems have to be considered when estimating uncertainties in abundance predictions. In the following, uncertainites in stellar reaction rates and in nucleosynthesis models are discussed. In particular, the impact of a measurement on the uncertainty of a stellar rate is scrutinized. Finally, numerical problems in stellar nucleosynthesis models are addressed in an example for the production of p-nuclei.

\section{DEFINING UNCERTAINTIES OF STELLAR REACTION RATES}
\label{sec:stellcs}

The stellar reaction rate (in reactions per time) for a reaction $a+A\rightarrow b+B$ in a plasma of temperature $T$ can be expressed as \cite{raureview,fow74,hwfz}
\begin{equation}
r^*(T) =  n_a n_A \sqrt{\frac{8}{\pi m_{aA}}} \left(kT\right)^{-3/2} \int_0^\infty \sigma_{aA\rightarrow bB}^*(E,T) E e^{-E/(kT)}\,dE =  n_a n_A \left< \sigma^* v \right>_{aA\rightarrow bB} = n_a n_A R^*(T)\quad,
\label{eq:stellrate}
\end{equation}
with the number densities $n_a$, $n_A$, the reduced mass $m_{aA}$ and Boltzmann constant $k$. The above equation also defines the reactivity $R^*=\left< \sigma^* v \right>_{aA\rightarrow bB}$ which is often used synonymously with the term ``rate''. The stellar cross section $\sigma^*(E,T)$ does not only depend on interaction energy but also on the plasma temperature which determines the population of excited states in the target. Explicitly showing the transitions between states $i$ in nucleus $A$ and final states $j$ in $B$, it can be written as \cite{raureview}
\begin{equation}
\sigma^*(E,T)=\frac{1}{G(T)} \sum_i \sum_j (2J_i+1) \frac{E-E_i}{E}
\sigma^{i \rightarrow j}(E-E_i) = \frac{1}{G(T)} \sum_i \sum_j W_i \sigma^{i \rightarrow j}(E-E_i)
\quad
\label{eq:stellcs}
\end{equation}
The partial cross sections $\sigma^{i \rightarrow j}$ are evaluated at the energy $E-E_i$, noting the convention by \cite{fow74} that cross sections at zero or negative energy are vanishing. Target levels are characterized by their spin $J_i$ and excitation energy $E_i$, with the g.s.\ having $i=0$ and $E_0=0$ MeV. The nuclear partition function of the target is 
\begin{equation}
G(T)=\sum_i \left( 2J_i+1\right) e^{-E_i/(kT)} = \sum_i P_i \quad.
\end{equation}

It is often overlooked that the relative importance of transitions from excited states is given by the weights
\begin{equation}
W_i=\frac{E-E_i}{E}=1-\frac{E_i}{E} \quad,
\end{equation}
which show a \textit{linear} dependence on excitation energy, despite of the exponential decline of the thermal Boltzmann population factors $P_i$. Relying solely on the $P_i$ as indication of the importance of thermally excited states will strongly underestimate their contribution.

The importance of excited target state transitions can be quantified by the g.s.\ contribution to the stellar rate, which is defined as \cite{Xs}
\begin{equation}
\label{eq:gsxfactor}
X_0(T)=\frac{2J_0+1}{G(T)} \frac{\int\sigma_0(E)E e^{-E/(kT)}dE}{\int\sigma^*(E,T) E e^{-E/(kT)} dE} = \frac{2J_0+1}{G(T)} \frac{R_0}{R^*} \quad,
\end{equation}
with $\sigma_0(E)=\sum_j \sigma^{0\rightarrow j}$ being the g.s.\ reaction cross section.  It always holds that $0\leq X_0\leq 1$.
It is very important to note that this is different from the simple ratio $R_0/R^*$ (the so-called stellar enhancement factor) of g.s.\ and stellar reactivities, respectively, which has been used mistakenly in the past to quantify excited state contributions. Even at s-process temperatures, g.s.\ contributions can already be small, especially in the rare-earth region. The g.s.\ contributions also decrease strongly with increasing temperature. The only exceptions are reactions close to magic numbers which retain larger g.s.\ contributions also at higher $T$ owing to the larger level spacing. Regarding charged particle reactions, $X_0$ are even smaller in most cases, due to the shift of effective interaction energy to higher energies in the integration for the rate \cite{energywindows}. All $X_0$ are given in \cite{sensis}.

Standard experiments determine only $\sigma_0$ and thus cannot completely remove theoretical uncertainties from the stellar rate unless $X_0=1$. 
The $X_0$ can be used to combine a measured $\sigma_0$ and a theoretical rate to an improved stellar rate \cite{sensis,Xs}. To obtain an improved stellar reactivity $R^*_\mathrm{new}(T)=f^*(T)R^*(T)$, the theoretical stellar reactivity $R^*$ has to be corrected by a factor $f^*$ accounting for the experimental information on the g.s.\ reactivity $R^\mathrm{exp}_0$. The factor is given by
\begin{equation}
f^*(T)=1+X_0(T)\left(\frac{R_0^\mathrm{exp}}{R_0^\mathrm{theo}}-1\right) \quad.\label{eq:renormstellar}
\end{equation}
Obviously, also the remaining uncertainty of the stellar reactivity is affected by this procedure. Thus, the $X_0$ can also be used to calculate a new uncertainty of the combined, improved stellar rate \cite{advances}.  The new uncertainty factor $u^*_\mathrm{new}$ of $R^*_\mathrm{new}$ is constructed from the original (theory) uncertainty factor $u^*>1$ and the experimental uncertainty factor $U_\mathrm{exp}>1$ of $R^\mathrm{exp}_0$ applying\footnote{It can be trivially assumed that $u^*>U_\mathrm{exp}$, otherwise no improvement would be possible.}
\begin{equation}
u^*_\mathrm{new}(T)=U_{\mathrm{exp}}+(u^*(T)-U_{\mathrm{exp}})(1-X_0(T)) \label{eq:uncertainty} \quad.
\end{equation}
It is straightforward to see from Eq.\ (\ref{eq:uncertainty}) that the rate uncertainty will not be affected by the measurement if $X_0$ is small. It should be noted, however, that the above equation assumes that the uncertainties in the predictions of g.s.\ and excited state transitions are uncorrelated.

As a rule of thumb, when comparing g.s.\ contributions in forward reaction $a+A\rightarrow b+B$ and reverse reaction $b+B \rightarrow a+A$, the reaction direction with positive reaction $Q$-value almost always shows larger $X_0$. Among the comparatively few exceptions (due to the Coulomb suppression effect of excited state transitions \cite{coulombsupp}) are capture reactions, which always have much larger $X_0$ than their photodisintegration counterparts. In fact, $X_0<5\times 10^{-5}$ for astrophysically relevant photodisintegration rates \cite{raureview,sensis}. This is important in the context of the origin of p-nuclei in the $\gamma$-process, as it shows that photodisintegration experiments cannot provide a constraint of the \textit{stellar} photodisintegration rates appearing in the nucleosynthesis of p-nuclei. A detailed discussion of the problematic application of photodisintegration experiments to astrophysical topics is found in Section III.C.1 of \cite{advances}.

\section{ASTROPHYSICAL MODEL UNCERTAINTIES}
\label{sec:astro}

\begin{figure}
\includegraphics[width=0.98\columnwidth]{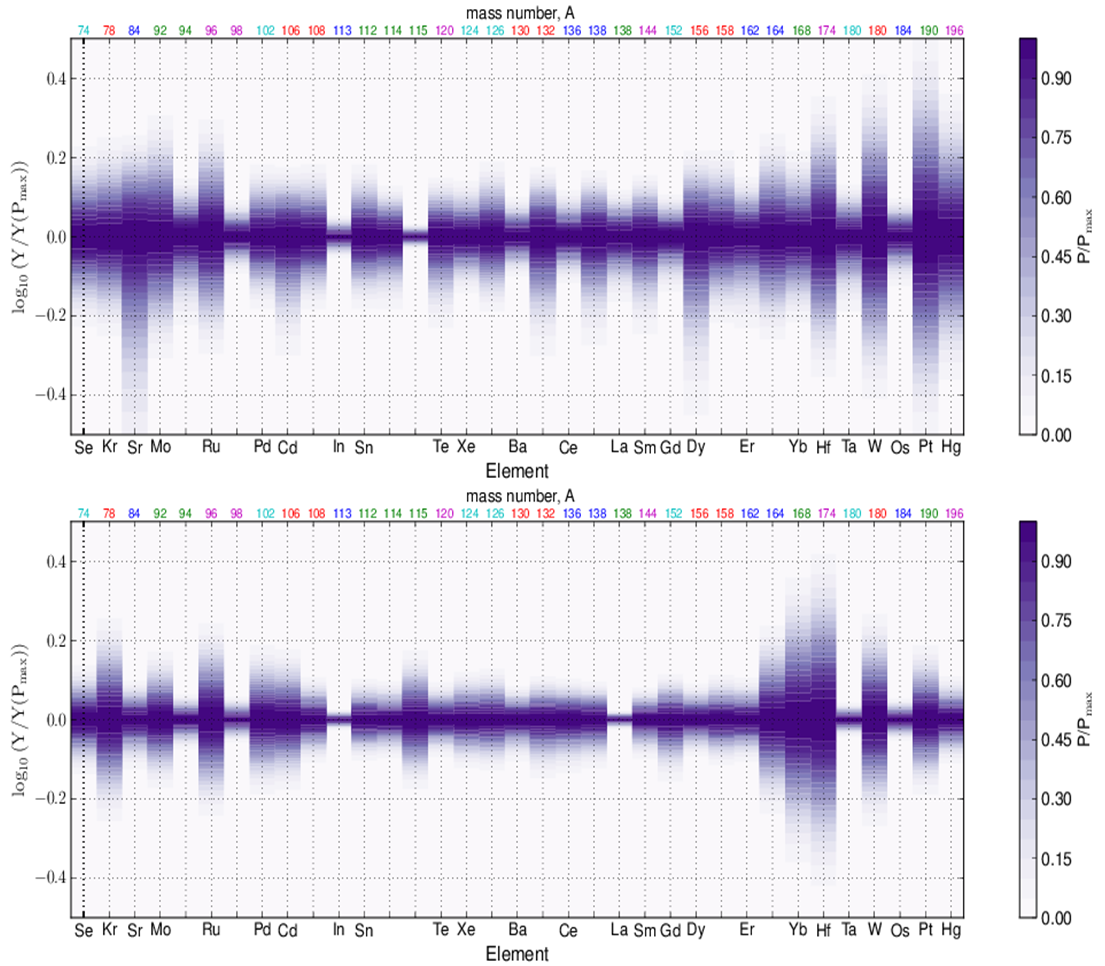}
\caption{Total uncertainties in the final abundances of each p-nucleus in the explosion of a 25 $M_\odot$ solar metallicity star, obtained with trajectories from \cite{hashi} (top) and \cite{rau02}. Shown are the normalized probability density distributions \label{fig:PDFs}}
\end{figure}

\begin{figure}
\includegraphics[width=\columnwidth]{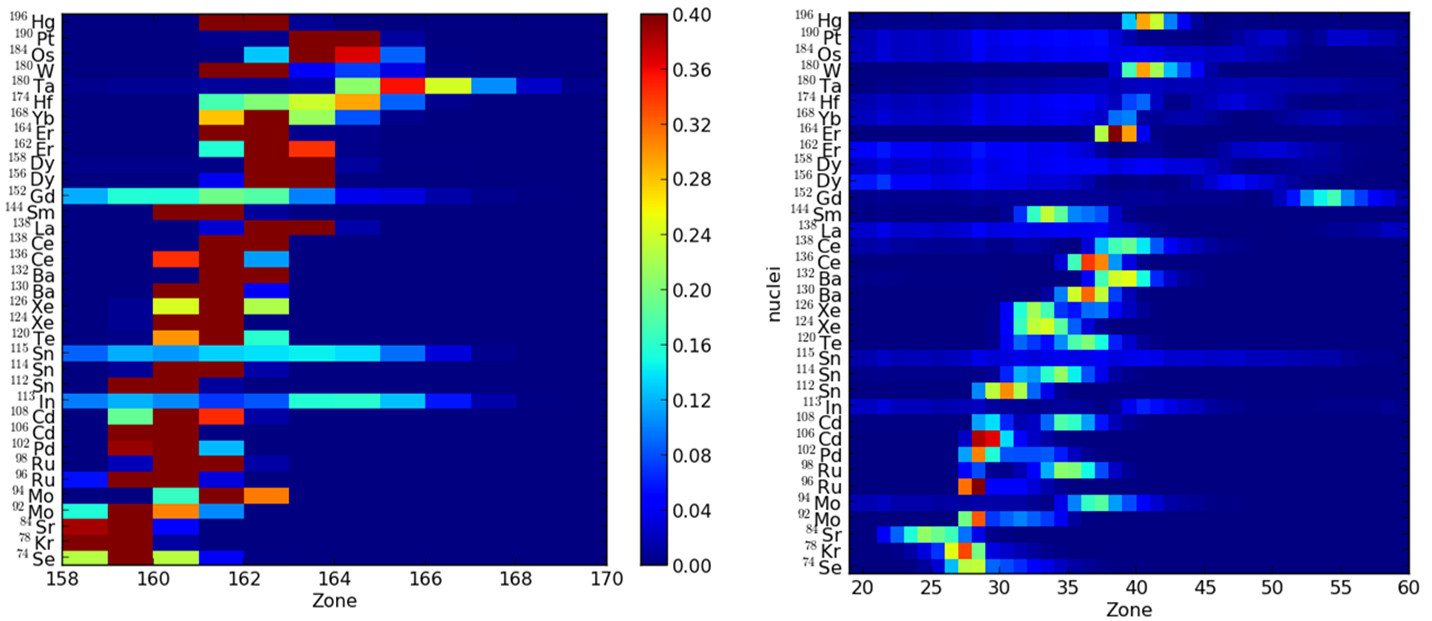}
\caption{\label{fig:zones} Relative abundance change for each p-nucleus in each zone. The model by \cite{hashi} (left) has larger zones, while the model by \cite{rau02} has a finer grid. The (arbitrary) zone number is given on the abscissa; lower zone numbers are deeper inside the star.}
\end{figure}

In addition to experimental and theoretical uncertainties in the determination of stellar reaction rates, uncertainties in astrophysical models have to be considered. Obviously, this includes the uncertainties in choosing the right representation of the actual physical processes leading to observed abundances. But this also includes the variations obtained from different numerical implementations of a given process. We illustrate this by an example for studying the origin of p-nuclei.

A small number (originally 35 but only 30 in today's literature \cite{dillmann}, as s- and r-process contributions were shown to be important for some of them) of proton-rich nuclei cannot be made in either s- and r-process and require other production mechanisms \cite{p-review}. For most of these so-called p-nuclei it has been shown that they can be made by photodisintegration of pre-existing nuclei in the outer layers of a massive star before and during its explosion as a core-collapse supernova (ccSN). The lightest of these nuclides, however, cannot be produced in sufficient quantities in this $\gamma$-process in massive stars and call for another explanation. Various suggestions have been made, such as a $\gamma$-process in type Ia supernovae, a $\nu$p-process in the deep layers of a ccSN, or a rp-process on the surface of mass-accreting neutron stars (see \cite{p-review} for an overview).

A detailed postprocessing analysis of p-nucleosynthesis has been performed by \cite{rayet95}, based on stellar models from \cite{hashi}. The trajectories from this model have been used subsequently for further studies and are also the basis for current studies to determine important reactions as possible targets for further experimental investigations. For example, a study collectively varying groups of reactions and using these trajectories was published in \cite{rapp}. We extended this study by performing a full-scale Monte Carlo (MC) variation using the PizBuin framework \cite{pizbuin,nic14}, varying reactions above the Fe group, including $\beta$ decays. Temperature-dependent variation limits have been used to account for the different uncertainties for experimentally determined g.s.\ contributions to the stellar rates and only theoretically determined excited state contributions (see above and \cite{advances,Xs}). Different uncertainty limits were assigned to different reaction types, with the temperature dependence obtained from Eq.\ (\ref{eq:uncertainty}). Experimental uncertainties were considered for g.s.\ contributions when available. In particular, predicted rates for neutron-induced reactions received an uncertainty limit of a factor of 2 (0.5), such involving protons a factor of 3.0 (0.33), whereas an asymmetric uncertainty was used for predicted rates with $\alpha$ particles. In the latter case, an uncertainty factor of 2.0 was assumed for the upper limit and a factor of 0.1 for the lower limit \cite{nic14}.
The actual MC variation factors for the rates were random values drawn from a uniform distribution between upper and lower uncertainty limit. Further details will be presented in a forthcoming paper.

Of interest here is that the method was not only applied with the trajectories for the explosion of a 25 $M_\odot$ star of solar metallicity by \cite{hashi} but also with those of the more recent model using the KEPLER code \cite{rau02}. In \cite{rau02}, the synthesis of p-nuclei was self-consistently followed in a large reaction network before and during the explosion of the star but for the MC study, trajectories and initial abundances were extracted from the zones offering favorable conditions for the $\gamma$ process and used for postprocessing.

Figure \ref{fig:PDFs} shows a comparison of the total uncertainty in the final abundance of each p-nucleus for the two models. The probability density distributions are encoded in color shade for each nuclide.\footnote{The nuclei $^{138}$La and $^{180}$Ta are mainly produced in the $\nu$-process which is not included in our postprocessing; the shown uncertainties only refer to other reactions involving these isotopes.} Although the general uncertainty patterns look similar, there are considerable differences for several nuclei in the results from the two models, which is somewhat surprising given that similar reaction flows should be contributing. Mainly responsible for these differences are the different zoning grids in the two models and the selection of zones. Figure \ref{fig:zones} shows the relative production (or destruction) of a given p-nucleus in each zone. The zoning in the model of \cite{hashi} is cruder than the one in \cite{rau02}, each zone encloses more mass and extends over a larger spatial distance. The temperature and density evolution in each zone is given by a separate trajectory. Zones deeper inside the star attain higher peak temperature in the passage of the supernova shock. A coarser grid, therefore, samples a coarser distribution of peak temperatures. Since the production and destruction of nuclei in the $\gamma$-process depends on the attained peak temperature, not all relevant reactions and reaction flows may by captured with a coarse zoning. This is especially true for the inner zones, for which the peak temperatures change more rapidly when moving from one zone to the next. Moreover, a coarse grid tends to overemphasize certain zones when a finer grid would actually find a significant change in abundance in several zones. In the context of uncertainties, this implies that certain reaction flows get more weight than they actually should, which results in a tendency to overestimate the uncertainty due to the lack of alternative paths.

A further problem can be identified in Fig.\ \ref{fig:zones}: The considered zones may not contain sufficient zones to completely follow the synthesis of light p-nuclei, as the zone cutoff lies too far out and several more zones further inside the star may still contribute. This not only impacts the resulting uncertainties but also directly the final abundances. It is clearly seen, on the other hand, that a sufficient number of zones from the KEPLER model was used to assure the inclusion of all zones relevant for p-nucleosynthesis. This is important to note, since the same set of trajectories from \cite{hashi} has been and is currently frequently used for postprocessing studies.

A similar problem was found in a recent comparison of p-nucleosynthesis in canonical thermonuclear supernovae.
For quite some time, an underproduction of light p-nuclei was found in such models \cite{p-review}. More recently, post-processing studies of high-resolution 2D models find production of all p-nuclei \cite{travWD}. It was concluded in \cite{kusa11,travWD} that a high-resolution treatment of the outer zones of the type Ia supernova is crucial to accurately follow the production of p-nuclides.

Regarding the $\gamma$-process in massive stars, it has to be noted further that stars across a range of masses contribute to the total budget of p-nuclei in the Galaxy. Due to differences in stellar structure and explosion energy, they produce different amounts and distributions of p-nuclei, as shown, for example, in \cite{rau02}. In a forthcoming publication we will investigate uncertainties in yields of p-nuclei from stars in a range of masses, to be used in galactic chemical evolution.

\section{CONCLUSION}
\label{sec:conclusion}

Many more transitions have to be known to determine stellar reactivities for trans-iron nucleosynthesis than in light element nucleosynthesis. This requires different theoretical and experimental approaches. Due to pronounced thermal plasma effects in the stellar rates, most measurements can only support predictions by testing models for single transitions but not constrain a stellar rate independently, especially in p-nucleosynthesis, with few exceptions. It has to be made sure that astrophysically relevant transitions are studied, though. Published systematic sensitivity studies and g.s.\ contributions to the stellar rate help to guide experiments. On the other hand, predictions are simplified by being able to average over many transitions and apply the Hauser-Feshbach model, which has been successful in describing a large number of reaction cross sections.

Astrophysical uncertainties not only include possible misrepresentations of the physics underlying a given observation but it has to be realized that also different model implementations and numerical differences are important. We have shown that a widely used set of trajectories for the $\gamma$-process in massive stars may be inadequate to follow the production of p-nuclei in detail.

\section{ACKNOWLEDGMENTS}
This work was partially supported by the European Research Council (grants GA 321263-FISH and EU-FP7-ERC-2012-St Grant 306901) and the UK Science and Technology Facilities Council (grant ST/M000958/1).
Numerical computations were in part carried out on the COSMOS (STFC DiRAC Facility)
at DAMTP at the University of Cambridge.

\end{document}